\documentclass[twocolumn,showpacs,preprintnumbers,amsmath,amssymb,prd]{revtex4}
\usepackage{graphicx}
\usepackage{dcolumn}
\usepackage{bm}

\begin{document}

\title{Are there indications of compositeness of leptons and quarks in CERN LEP data?}

\author{Erik Elfgren}
 \email{elf@ltu.se}
\author{Sverker Fredriksson}
 \email{sverker@ltu.se}
\affiliation{%
Department of Physics\\
Lule\aa \ University of Technology\\
SE-97187 Lule\aa, Sweden}


\begin{abstract}
The ``preon-trinity'' model for the
compositeness of leptons, quarks and heavy
vector bosons predicts several new heavy leptons
and quarks. Three of them can be produced
in $e^{+}e^{-}$ annihilations at CERN LEP
energies, since they can be created
out of a system of three preons and their
antipreons, where three preons form
a heavy lepton or quark, while the other three
go into a normal lepton or quark.
In fact, these new particles are predicted
to be lighter than the top quark,
while the top itself cannot be produced this
way, due to its particular preon substructure.
The empirical situation is analyzed, and the most likely
masses are estimated.
\end{abstract}


\maketitle

\section{Introduction}
New generations of leptons and quarks,
as compared with those prescribed by
the so-called standard model (SM), have been searched for
at the three main high-energy laboratories,
{\it i.e.}, CERN LEP, DESY HERA and
the Fermilab Tevatron. A general conclusion
is that no statistically significant
signals have been found \cite{pdg}.
This goes for both a fourth generation
of leptons and quarks, {\it e.g.}, a $b'$
quark, and excited versions of the normal
ones, {\it e.g.}, $e^{*}$ and $\nu^{*}$.
From an experimental point of view there
is not much difference between the two
approaches. A fourth generation is believed to
mimic the normal three, while excited leptons
and quarks are believed to couple
to other particles exactly like the
unexcited versions, except for
kinematic effects of the higher masses.

The existence of any of these would be a
strong evidence for a substructure of
leptons and quarks in terms of preons.
Excitations are hard to imagine without
an inner structure of the excited object,
and yet another generation of leptons and quarks
would also be difficult to reconcile
with the idea that they are all fundamental.

There are indeed already several phenomenological,
and logical, arguments in favour of a
substructure of leptons, quarks and heavy
vector mesons in terms of preons
\cite{souza,fredriksson,kalman}. It even
seems as if the standard model itself
contains several prophecies of preons
\cite{fredriksson} among its many seemingly
unrelated bits and pieces.
The ``preon-trinity'' model \cite{dugne}
was inspired by such arguments,
as well as by models for
the three-quark structure of light baryons
\cite{gellmann,zweig}, by
early preon models \cite{souza,harari,shupe,fritzsch}
and by the concept of diquarks \cite{anselmino}.

The aim of this publication is to reanalyze the data from,
above all, the (closed) CERN LEP facility, in order to look
for signals of new leptons and quarks, as prescribed
by the preon-trinity model. As it turns
out, some of the criteria used in the existing experimental
searches are not valid in the model. Above all, the
predicted new leptons and quarks are {\it not} just
heavier versions of the old ones. They have their own
unique features, which should be confronted with the
data.

\section{The Model}
The main ingredients are that there
exist three absolutely stable species
(flavours) of spin-$1/2$ preons, called $\alpha$,
$\beta$ and $\delta$, with electric and
colour charges, and that these also
tend to form tightly bound spin-$0$ dipreons.
Thanks to the choice of preon charges, inspired
by the ones of the original three-quark model,
the scheme gets an attractive
supersymmetric balance between preons and
{\it anti-}dipreons, as summarised in Table I.

\begin{table}
\begin{center}
\caption{The ``supersymmetric'' preon scheme.}
\vskip0.1cm
\begin{tabular}{l|ccc}
charge & $+e/3$ & $-2e/3$ & $+e/3$\\
\hline
spin-$1/2$ preons & $\alpha $ & $\beta $ & $\delta $ \\
spin-$0$ (anti-)dipreons & $(\bar{\beta} \bar{\delta})$ &
$(\bar{\alpha} \bar{\delta})$ &
$(\bar{\alpha} \bar{\beta})$\\
\end{tabular}
\end{center}
\end{table}
It is then prescribed that leptons are built up
by one preon and one dipreon. Quarks consist
of one preon and one anti-dipreon, and heavy
vector bosons of one preon and one antipreon.
The results are shown in Table II. There is an
obvious $SU(3)$ preon-flavour symmetry in
the scheme, just like with quark flavours in the
first quark model. The stability of preons
means that the total preon flavour is absolutely
conserved, unlike the quark flavour in the quark
model.

\begin{table}
\begin{center}
\caption{Composite states in the preon model:
{\it leptons} as a preon and a dipreon,
{\it quarks} as a preon and an anti-dipreon,
and {\it heavy vector bosons} as a preon and an antipreon.}
\vskip0.3cm
\begin{tabular}{c|ccc|ccc|ccc|}
& $(\beta \delta)$
& $(\alpha \delta)$
& $(\alpha \beta)$
& $(\bar{\beta} \bar{\delta})$
& $(\bar{\alpha} \bar{\delta})$
& $(\bar{\alpha} \bar{\beta})$
& $\bar{\alpha}$
& $\bar{\beta}$
& $\bar{\delta}$  \\
\hline
$\alpha$
& $\nu_{e}$
& $\mu^{+}$
& $\nu_{\tau}$
& $u$
& $s$
& $c$
& $Z^{0}/Z'$
& $W^{+}$
& $Z^{*}$ \\
$\beta$
& $e^{-}$
& $\bar{\nu}_{\mu}$
& $\tau^{-}$
& $d$
& $X$
& $b$
& $W^{-}$
& $Z'/Z^{0}$
& $W'^{-}$ \\
$\delta$
& $\nu_{\kappa1}$
& $\kappa^{+}$
& $\nu_{\kappa2}$
& $h$
& $k$
& $t$
& $\bar{Z}^{*}$
& $W'^{+}$
& $Z''/Z'$   \\
\end{tabular}
\end{center}
\end{table}
One can make about a dozen observations about
leptons, quarks and heavy vector bosons. Most of
these provide qualitative explanations
of some seemingly disjunct ingredients of the SM,
{\it e.g.}, the mixings of some quarks,
neutrinos and heavy vector bosons, and
the (partial) conservation of three lepton
numbers, all being consequences of preon
flavour conservation \cite{dugne}.

The most notable difference from the SM
is the set of new leptons, quarks
and heavy vector bosons predicted by the model.
These contain a $\delta$ preon that does
not belong to a dipreon, and are to be found in
the bottom row of Table II, plus in the right-most
column. They must all have such high
masses that they have escaped discovery,
{\it except} the $t$ quark, which is most
probably the quark below $b$ in Table II. This
indicates that also the other new quarks and leptons
have masses in the region $100-200$~GeV.
There is a good chance that $\nu_{\kappa1}$,
$\kappa$, $\nu_{\kappa2}$, $h$ and $k$
(earlier called ``$g$'' \cite{dugne})
are all {\it lighter} than the top quark.
It must be stressed though that the model is entirely
defined by this scheme, and has not (yet) been
complemented with a preon dynamics.
Hence exact masses, branching ratios and
life-times of quarks and leptons
cannot be reproduced or predicted.
However, experience tells that lepton masses are
lower than that of ``corresponding'' quarks. In addition,
there is a trend among the known quarks of Table~II that
the masses increase from left to right. All in all, one
can therefore expect the mass relations
$M_{\nu_{\kappa1}} < M_{\nu_{\kappa2}} < M_{t}$, $M_{h} < M_{k} < M_{t}$, and $M_{\kappa} < M_{k}$,
which means that CERN LEP energies would suffice for
the reactions discussed in this study, and that signs of
compositeness might hence exist in old data.

In the following, only the relevance of the model to
existing data from the CERN LEP facility will
be discussed. For other details of the model,
such as the argument that the odd $X$ quark,
with charge $-4e/3$, is not expected to exist
as a bound system, the reader is referred to
Ref. \cite{dugne}. General arguments
that preons should exist are given in
Refs. \cite{souza,fredriksson}.

\section{The Reactions}
Any system of a certain preon flavour and its
anti-flavour can be produced in
$e^{+}e^{-}$ annihilation, as long as the
energy suffices. As can be seen in Table II,
there are three new leptons/quarks that carry
the same net preon flavour as a lighter
partner. These pairs are
$\nu_{\kappa2} = \delta(\alpha\beta)
\leftrightarrow \nu_{e} = \alpha(\beta\delta)
\leftrightarrow \bar{\nu}_{\mu} = \beta(\alpha\delta)$,
$h = \delta (\bar{\beta}\bar{\delta})
\leftrightarrow c = \alpha (\bar{\alpha}\bar{\beta})$
and
$k = \delta (\bar{\alpha}\bar{\delta})
\leftrightarrow b = \beta (\bar{\alpha}\bar{\beta})$.

As long as the masses of these leptons and quarks 
are not too high, $e^{+}e^{-}$ annihilation can
therefore result in the production of the pairs
$\nu_{\kappa2}\bar{\nu}_{e}$,
$\nu_{\kappa2}\nu_{\mu}$,
$h\bar{c}$ and $k\bar{b}$:
\begin{equation}
e^{+}e^{-} \rightarrow
\delta(\alpha\beta)
+ \bar{\alpha}(\bar{\beta}\bar{\delta})
= \nu_{\kappa2} + \bar{\nu}_{e},
\end{equation}
\begin{equation}
e^{+}e^{-} \rightarrow
\delta(\alpha\beta)
+ \bar{\beta}(\bar{\alpha}\bar{\delta})
= \nu_{\kappa2} + \nu_{\mu},
\end{equation}
\begin{equation}
e^{+}e^{-} \rightarrow
\delta (\bar{\beta} \bar{\delta})
+ \bar{\alpha} (\alpha \beta)
= h + \bar{c},
\end{equation}
and
\begin{equation}
e^{+}e^{-} \rightarrow
\delta (\bar{\alpha} \bar{\delta})
+ \bar{\beta}(\alpha \beta)
= k + \bar{b}.
\end{equation}
(and the corresponding antiparticles). For the sake
of simplicity, reactions with additional particles
in the final state, {\it e.g.}, photons,
will not be discussed here. All final particles
will hence be assumed to come from hadronisation
or decay of the quarks or leptons listed above.

It is notable that the top quark,
$t = \delta(\bar{\alpha}\bar{\beta})$, has no
such partner, and hence there is no
``single-top production'' in $e^{+}e^{-}$
reactions.

Since $e^{-} = \beta(\beta\delta)$, {\it all}
processes of interest are transitions from the
original preon system
$\beta(\beta\delta)\bar{\beta}(\bar{\beta}\bar{\delta})$
to the intermediate one
$\alpha\bar{\alpha}\beta\bar{\beta}\delta\bar{\delta}$.
This, in turn, can split up in many different ways,
among which are the four different pairs listed
above.

The transition
\begin{equation}
e^{+}e^{-} =
\beta(\beta\delta)\bar{\beta}(\bar{\beta}\bar{\delta})
\rightarrow S^{*} \rightarrow
\alpha\bar{\alpha}\beta\bar{\beta}\delta\bar{\delta}
\end{equation}
can take place via a number of intermediate systems $S^{*}$.
Examples are one or more photons, a suitable
combination of gluons (or ``hypergluons'';
the possible quantas of a hypothetical
preon interaction \cite{souza}), a preon-antipreon pair
($\beta\bar{\beta}$, after the annihilation of
$(\beta\delta)(\bar{\beta}\bar{\delta})$) or
a dipreon-anti-dipreon pair
($(\beta\delta)(\bar{\beta}\bar{\delta})$, after
the annihilation of $\beta\bar{\beta}$).

The latter should be less likely, because it seems
as if a dipreon always stays together, once it has
been created \cite{dugne}. Hence the only new
particles that can be produced from the annihilation
of $\beta\bar{\beta}$ would be through
the creation of a $\delta\bar{\delta}$ pair.
This would result in either
$e^+e^- \rightarrow \nu_{\kappa1}\bar{\nu}_{\kappa1}$
or $h\bar{h}$, with only superheavy final leptons or quarks.
They can occur only if $M_{\kappa1},M_h< E_{LEP}/2$,
where $E_{LEP}$ is the total LEP energy
($\leq 209$~GeV).
In that case they might be seen through the decay
products of the two neutrinos or quarks.

The model also allows for production processes like
\begin{equation}
e^{+}e^{-} \rightarrow \nu_{e} + \nu_{\mu}
\end{equation}
(but {\it not} of
different {\it charged} leptons, like $e^{+}\mu^{-}$).
However, due to the neutrino helicities, this can happen
only for annihilation in a total spin-$0$ state.
Therefore, this final state cannot be produced in,
{\it e.g.}, $Z^{0}$ decay, which means that it cannot
be restricted by the well-known
``three-generation'' data from $Z^{0}$ decay
\cite{pdg}. However, the decay
\begin{equation}
Z^{0} \rightarrow \bar{\nu}_{e} + \nu_{\kappa 2},
\end{equation}
should, in principle, be possible,
and similarly for the final states with quarks in
reactions (3) \& (4). They have not been seen,
so their masses must exceed $M_Z/2$.

\section{Analyzing the CERN LEP data}
We will start by discussing the $\nu_{\kappa 2}$ decay channels of
interest for an analysis of LEP data. It is important to keep
in mind that lepton numbers are not exactly conserved
in our model, and that there is no fourth lepton number
connected to the predicted new leptons. The observed
lepton number conservation is in our model equivalent
to ``dipreon number conservation'', {\it i.e.}, the three usual
lepton numbers are conserved in leptonic processes
only to the extent that the tightly bound dipreons are left intact.
In normal leptonic decays and ``low-energy'' reactions
this must be the case,
because all imaginable dipreon-breaking processes
violate energy conservation. However, the heavy
leptons {\it must} decay through a reshuffling of the
preons inside a dipreon, and would hence change the
normal lepton numbers. An example is
$\kappa^{+} \rightarrow \mu^{+} + \nu_e + \bar{\nu}_{\tau}$,
violating all three lepton numbers.
In addition, the
three neutrinos on the diagonal of Table II might mix
into new mass eigenstates, since they have identical
preon net flavours. This is equivalent to neutrino
oscillations, which do not conserve lepton numbers.
Lepton number conservation might also be violated in
normal leptonic collisions, if the energy is high enough to
break up existing dipreons. As argued in \cite{dugne}
we think that the energy scale for new preon
processes is a few hundred GeV rather than TeV
(as the top quark seems to be an example
of a ``superheavy'' three-preon state). The
``TeV scale'' often mentioned in discussions
of compositeness is rather the expected momentum-transfer
scale for revealing substructure in deep-inelastic
scattering of leptons and quarks.

The most interesting decay channel for the
lightest heavy neutrino is
\begin{equation}
\nu_{\kappa 2} \rightarrow e^{-}/\mu^{+} + W^{+/-},
\end{equation}
followed by
\begin{equation}
W \rightarrow q_{1}\bar{q}_{2},
\end{equation}
or in terms of preon processes:
\begin{equation}
\delta (\alpha \beta) \rightarrow
\beta (\beta \delta)/ \alpha (\alpha \delta)
+ (\alpha \bar{\beta})/(\bar{\alpha} \beta).
\end{equation}

The decay into two quark jets
gives an opportunity to find the invariant
mass of the neutrino. The $W$ is most probably real if
the neutrino mass is reasonably well above
the $W$ mass. Hence one should restrict
the analysis to events where the estimated
invariant mass of the two hadron jets is close
to the $W$ mass. The main background to this process is
$e^{+}e^{-} \rightarrow W^{+}W^{-}$ followed by one
$W \rightarrow \ell + \bar{\nu}$ and the other
$W \rightarrow q_{1}\bar{q}_{2}$, where $\ell$ is $e$ or $\mu$.

The analysis hence requires a standard-model
Monte Carlo simulation, where one looks for an excess of
events within an interval of invariant masses around
some value below $175$~GeV (the top mass).
If such an excess is seen, the ``extra events'' should have
a few characteristics, typical for our model, but not for
the standard-model background events:

\begin{itemize}
\item there would be a threshold effect at the total
$e^{+}e^{-}$ energy $\sqrt{s} = M_{\nu_{\kappa 2}}$,
rather than at the $W^{+}W^{-}$ threshold.
\item the charged lepton would always be back-to-back to the $W$
({\it i.e.}, the cms of the two hadron jets) in the rest system of
the $\nu_{\kappa 2}$. This, in turn, has a speed given by
kinematics only, {\it i.e.}, by $\sqrt{s}$ and $M_{\nu_{\kappa 2}}$.
\item at first sight it seems as if the
$\nu_{\kappa 2}$ would decay as willingly to an electron
as to a muon in accordance with, {\it e.g.},
$W$ decay. However, the situation is a bit more
complicated, since $e^{-}$ and $\mu^{+}$
(not $\mu^{-}$) are on equal footing in our model.
They have {\it opposite} dominant helicity
components in the ultra-relativistic limit. Although
we do not know the dynamics of the heavy-neutrino
decay, we suspect that its decay favours an outgoing
charged lepton with positive helicity, {\it i.e.},
the $\mu^{+}$ (and $\mu^{-}$ in $\bar{\nu}_{\kappa 2}$
decay). Intuitively, this seems in accordance
with the helicities of the
decay $W^{+} \rightarrow \mu^{+} + \nu_{\mu}$.
\end{itemize}

These predictions are best investigated by the
CERN OPAL collaboration. It has so far published
searches for, among others, heavy neutral leptons
at $\sqrt{s}$ values up to $183$ GeV
\cite{OPAL96,OPAL98,OPAL00}. A similar analysis at
the highest LEP energies is underway \cite{plane}.

The main conclusion so far from OPAL is that no
signs of a heavy neutral lepton have been found at
LEP energies up to $183$ GeV. This result is summarised
as $95\%$ CL lower mass limits for various
channels (decay modes), the values being
typically around $90$ GeV.

However, the OPAL analyzes contain some extra
assumptions, not necessarily valid for
our model. For instance, a new heavy
neutrino is supposed to belong to a new
``fourth family'', and be produced
toghether with its own antineutrino. This is
not the case in our model. In the search
for excited versions of the normal neutrinos
it is supposed that the couplings are
as prescribed by the standard model.
Any differences between
a normal neutrino and its excited partner
is assumed to be due to the different masses only.
We therefore look
forward to a less model-dependent
analysis of {\it all} available OPAL
data along the directions outlined
above, including the most recent
ones beyond $200$ GeV.

Next we consider the decay channel
\begin{equation}
\nu_{\kappa 2} \rightarrow e/\mu + W,
\end{equation}
\noindent followed by
\begin{equation}
W \rightarrow e/\mu + \nu_{e}/\nu_{\mu}.
\end{equation}
Now the invariant neutrino mass cannot be
derived. The only signal of a new neutrino would
therefore be an excess of events compared to what
is expected from the standard model in the channel
\begin{equation}
e^{+}e^{-} \rightarrow
\ell_{1}^{+}\ell_{2}^{-} + invisibles.
\end{equation}

Here also the ALEPH collaboration can contribute
to the analysis, although it focuses on
$W$-pair production at various LEP energies
\cite{ALEPH97a,ALEPH97b,ALEPH99,ALEPH00,ALEPH01}.
This means that there are several experimental
cuts in order to assure that each event produces
a $W$-pair, which might naturally eliminate
alternative processes, such as the one we are
interested in.

Looking at the totality of ALEPH data,
for all $W$ decay modes, the ones taken at the lowest
three LEP energies \cite{ALEPH97a,ALEPH97b,ALEPH99}
are consistent with a small
``excess'' of events in some kinematic bins,
but only on the $0.5 \sigma$ to $1 \sigma$ level.
The case is weakened by the fact that other bins
show similar ``deficits'' in comparison to
Monte Carlo simulations of the standard model,
hinting at a mere statistical effect.
In addition, ALEPH does not find any relevant
deviation from lepton universality, {\it e.g.}, an excess
of muons, according to our speculation above
that a heavy neutrino would prefer muonic decays.

At the LEP energy of $189$ GeV the ALEPH statistics
is much better, and the possible deviations from
the standard model are even smaller, on the order
of $0.2 \sigma$ at the most.

The only aspect of the ALEPH data that can give
some support to a closer analysis along our prediction
is the fact that there is a clearer excess of events
in kinematic regions outside the (``acoplanarity'')
cuts used to define $W$-pair production, especially for LEP
energies up to $183$~GeV. Obviously, these regions are
expected to contain events with other configurations
than just a $W$ pair, but it is not clear to us
why the simulations do not describe the data so well.

We now analyze the different neutrino decay mode
\begin{equation}
\nu_{\kappa 2} \rightarrow \nu_{e}/\nu_{\mu}
+ \gamma (\gamma),
\end{equation}
or in terms of preon processes:
\begin{equation}
\delta (\alpha \beta) \rightarrow
\beta (\alpha \delta)/ \alpha (\beta \delta) + \gamma (\gamma).
\end{equation}
Hence the full events of interest are
\begin{equation}
e^{+}e^{-} \rightarrow \nu \bar{\nu} + \gamma (\gamma).
\end{equation}
The signal is one or more high-energy gammas,
and a deviation in the
production cross-section and the phase-space distribution of
gammas, as compared to expectations from the standard model.

Such studies have been made by the DELPHI collaboration
\cite{DELPHI99a,DELPHI99b,DELPHI00}, in events with just photons
(plus ``invisibles''), and at LEP energies up to
$189$ GeV. Parts of the DELPHI analysis focus on the
possibility of a new generation of heavy neutrinos.

The results are summarised as the distribution
in ``missing mass'' of the invisibles recoiling
against the gamma(s). The possible deviations
of this distribution from the standard-model
expectation are presented as upper production
cross-section limits of a heavy ``neutral object''.
However, this analysis is built on the idea that
the heavy neutrino is ``stable'', and hence
identical to one of the outgoing neutrinos
in the process given above. Provided that its mass
is fairly high, and that only high-energy gammas
are studied, the ``missing mass'' would be
a good measure of the neutrino mass.
A theoretical analysis of expected event rates
for such a production of a $50$ GeV neutrino
at LEP is presented in \cite{ilyin}.

If we stay with the case of a highly
unstable neutrino,
where the gammas come from its decay
(and not from its production), there is
a simple relation, in the one-gamma case,
between the ``missing mass'' ($MM$)
of the DELPHI analysis
and the mass ($M_{\kappa2}$)
of $\nu_{\kappa2}$. Assuming that
its decay products, $\gamma$ and a light
neutrino, are aligned along its spin direction,
we get:
\begin{equation}
M_{\nu_{\kappa 2}} = \sqrt{E_{LEP}^{2} - MM^{2}}.
\end{equation}
There is no such simple relation for two-gamma decays,
since we do not know the dynamics behind the decay.
Intuitively, it seems likely that the ``missing
mass'' distribution would anyway peak at the same value
as for the one-gamma decays, but there might
also be a second peak due to the fact that the
gammas can radiate in the same or in opposite
directions in the $\nu_{\kappa2}$ rest system.
The only data points that
deviate by more than $1\sigma$ from the standard-model
result in \cite{DELPHI00} are at $MM \approx 135$ GeV
in both the $2\gamma$ data and in parts of the
one-gamma data (from the HPC calorimeter). This
value corresponds to a $\nu_{\kappa2}$ mass of
around $140$~GeV. There is also a similar excess
at $MM \approx 165$ GeV in the $2\gamma$ data,
possibly corresponding to a $110$~GeV neutrino.
However, a signal would be smeared out, since
the data are summarised over several LEP energies.
No significant excess is seen in the one-gamma
case when the data are summed from all (three)
DELPHI calorimeters.

The detailed analysis of
``limits of compositeness'' in \cite{DELPHI00}
is not of much value for judging our ideas,
because it is built entirely on predictions
from a rather specific preon model \cite{senju}.
These model predictions rely, for instance, on
the existence of additional, composed bosons
with unknown (high) masses.

Finally, one might ask if the lightest of the
superheavy neutrinos, the $\nu_{\kappa1}$,
might be {\it stable}, and hence correspond to
the hypotherical ``dark-matter'' neutrino
analyzed in \cite{elfgren} and elsewhere.
Since it is most probably the lightest of
the superheavy particles, it can perhaps
be pairproduced at LEP as
$\nu_{\kappa1}\bar{\nu}_{\kappa1}$
(but it cannot be created
together with a light partner).
However, there is no particular reason for
a stable $\nu_{\kappa1}$. All superheavy
leptons and quarks {\it must} decay through
the break-up of their dipreons. Otherwise the
top quark would be stable, and so would the
heavy lepton $\kappa$. The $\nu_{\kappa1}$
would indeed decay to, for instance, an $e^-$
through the same preon processes as for
$t \rightarrow b$, as can be seen in
Table~II.

A new, heavy quark can be studied in two
different ways. One can either look for
an excess of events with a high missing mass
that recoils against the normal quark. Or one
can try to identify the decay products of
the new quark, and derive their invariant mass.

The first method should, in principle, be simpler
than the second one, since it takes only to identify
a $c$ or $b$ jet and measure its energy.
Assuming that no other particles have been
produced than a pair of one new and one
normal quark, and that the mass of the latter
can be neglected, one gets the relation
\begin{equation}
E_{q} = \frac{E_{LEP}^2 - M^2}{2E_{LEP}},
\end{equation}
between the mass $M$ of the new quark and the
energy $E_{q}$ of the recoiling normal quark.
In practice, jet energies are measured,
and the relation between the initial quark energy
and that of its final hadronic jet is not
simple. In order to search for the $h$ (or $k$), quark
data are needed for jets typical for $c$ (or $b$)
quarks. Unfortunately, most LEP data on heavy flavours
are taken at the $Z^0$ peak. Superheavy quarks would
give broad peaks in the jet energy
spectrum according to Eq. (18), on top of the
background, which would most probably come from
$e^+e^- \rightarrow W^+W^-$ followed by
$W \rightarrow q_1\bar{q}_2$.

The second method, {\it i.e.}, to look for
decay products of the heavy quark or lepton,
is the conventional one. The two quarks
$h$ and $k$ do not decay like the $t$,
and cannot be discovered as a result
of the search for single-top production.
However, $k$ is identical to the
hypothetical fourth-generation $b'$ quark,
which has been searched for in many
experiments \cite{pdg}. It decays like
$k \rightarrow b + x$,
through $\delta (\bar{\delta}\bar{\alpha}) \rightarrow
\beta (\bar{\beta}\bar{\alpha}) + x$,
where $x$ can be a $\gamma$, a $Z^{0}$ or a gluon.
According to \cite{oliveira}
the $bZ^0$ or $cW$ channels should dominate
the decay of a ``$b'$'', so one way to find it would be
to analyze the recoiling mass against a $b$ and
a $Z^0$ jet, or a $c$ and a $W$ jet.
However, their analysis is built on production
of $b'\bar{b}'$ {\it pairs}, unlike in our model.
And so is the recently published experimental
search by the DELPHI collaboration \cite{DELPHI07}.
We instead suggest that the analysis is
remade for $b'\bar{c}$ production, which would
require other kinematical cuts than in \cite{DELPHI07}.

In conclusion, we argue that there might still
be room for signatures of composite quarks
and leptons in the CERN LEP data, provided that
these are analyzed with slightly different
methods in comparison to what has been done
so far.

We acknowledge useful communications with
R.~Tenchini from the ALEPH collaboration,
J.~Timmermans and S.~Ask from the DELPHI collaboration,
as well as D.~Plane from the OPAL collaboration.
E.E. is grateful to the Swedish National Graduate
School of Space Technology for financial support,
and S.F. would like to thank the Max Planck Institut
f{\"u}r Kernphysik, CERN and Indiana University
for hospitality and stimulating feedback during visits
where these ideas were presented and discussed.

\end{document}